\documentclass{pasj00}
\draft

\begin{document}
\SetRunningHead{Y. Fujita}{Infant Galaxy Clusters?}
\Received{0000/00/00}%{yyyy/mm/dd}
\Accepted{0000/00/00}%{yyyy/mm/dd}

\title{Infant Galaxy Clusters at Low Redshifts?}

%%% begin:list of authors
\author{Yutaka \textsc{Fujita}\altaffilmark{1,2}}%
%  \thanks{Example: Present Address is xxxxxxxxxx}}
\altaffiltext{1}{National Astronomical Observatory, Osawa 2-21-1,
Mitaka, Tokyo 181-8588}
\altaffiltext{2}{Department of Astronomical Science, 
The Graduate University for Advanced Studies,\\ Osawa 2-21-1,
Mitaka, Tokyo 181-8588}
\email{yfujita@th.nao.ac.jp}

%% `\KeyWords{}' always has to be placed before `\maketitle'.
 %Do NOT move this preamble from here!

\maketitle

\begin{abstract}
The population and population composition of galaxies in galaxy clusters
at present reflect the mass of the clusters and the mass growth of the
galaxies in the past. We investigate them for six clusters. We show that
galaxies in massive clusters stopped growing at redshifts of $\sim
4$. Moreover, we find that some small galaxy clusters (groups) have too
many massive galaxies for their apparent masses. One possibility is that
these groups are much more massive and in a phase just before
virialization. If this is the case, they should be called `infant galaxy
clusters' that will be matured clusters in the dynamical time-scale
($\sim 10^9$ yrs).
\end{abstract}

\section{Introduction}

Clusters of galaxies are the most massive objects in the universe. It is
generally believed that dark matter constitutes a large fraction of the
mass in the universe, and galaxies and clusters form in dark
matter-dominated halos (dark halos). In the standard picture of
hierarchical structure formation of the universe, small objects are the
first to form and these then amalgamate into progressively larger system
\citep{blu84}. This means that a dark halo containing a galaxy cluster
has formed via merging and accretion of smaller dark halos containing
galaxies or galaxy groups (small galaxy clusters). From now on, we do
not discriminate a dark halo containing a cluster from the cluster
unless otherwise mentioned; the same rule is applied for a group and a
galaxy.

The mass distribution function (MDF) of galaxies absorbed into but not
disrupted in a cluster is defined by the number of the galaxies in a
given mass range and shows the population composition of the galaxies
with different masses. The MDF tells us the average time when the
galaxies stopped growing as follows. The mass of galaxies increases with
time as matter accretes to them through gravity. As a result, the number
of massive galaxies increases and the MDF evolves. However, as the
potential well of the host cluster (or its progenitors) deepens as the
cluster grows up, the released gravitational energy of the cluster
allows the galaxies to move fast in the cluster. The large velocity of
the galaxies prevents them from gathering more matter and the galaxies
cannot grow anymore. Thus, the MDF freezes after that. The MDFs observed
at present should depend on the time of the freeze. On the other hand,
the integrated MDF or the galaxy population in a cluster depends on the
host cluster mass; the more massive a host cluster is, the more massive
galaxies are contained in it.

We have devised an analytical model to calculate the MDFs of galaxies in
clusters \citep{fuj02a}. In the field of cosmological study, the
Press-Schechter (PS) model has been used to obtain an average MDF of
objects (galaxies and clusters of galaxies) in the universe
\citep{pre74}. The PS model treats the evolution of initial density
fluctuations of the universe and predicts the number of collapsed
objects in a give mass range. However, this model cannot be used to
obtain the MDFs of galaxies in clusters, because it does not take
account of two important effects unique to clusters. One is that even
before a cluster forms, the precluster region should have a higher
density than the average of the universe. The other is the spatial
correlation of initial density fluctuations of the universe. The
correlation is important when the fluctuations are close to each other;
a precluster region is an example of the crowded fluctuations.  The
former effect has already been involved in the so-called extended
Press-Schechter model (EPS) \citep{bow91,bon91,lac93}. The model used
here is a further extension of the EPS model by taking account of the
latter effect (the spatial correlation), so we refer to this model as
the SPS model. The details of the SPS model are described in
\citet{fuj02a}.

\section{Comparison between Theoretical Predictions and Observations}

In principal, we can calculate MDFs using the SPS model and compare them
with the observed ones to estimate redshifts when the observed MDFs
almost froze (that is, the formation redshifts of galaxies; $z_f$). We
note that $z_f$ is the {\it average} redshift when the growth of
galaxies are held by external force (that is, the growth of the host
cluster). At $z=z_f$ the host cluster had not necessarily been unified,
and the host cluster might be divided into several progenitors. We can
say that the redshift $z_f$ corresponds to the time when the {\it
average} mass of those progenitors became large enough to hold the
growth of most of the galaxies that are included in the host cluster at
$z=0$. Because it is not easy to measure galaxy masses directly by a
method like gravitational lensing, it would be difficult to derive MDFs
of galaxies in clusters observationally. Thus, we convert the MDFs
derived using the SPS model to velocity distribution functions (VDFs) of
galaxies by an empirical relation between the internal velocity and mass
of galaxies obtained by numerical simulations\footnote{The original
paper assumed that the relation between the mass and internal velocity
of galaxies is time-independent (equation~[40] in \cite{fuj02a}). In
this paper, we consider the time dependence as equation
(\ref{eq:vint}).}(Fig.~13 in \cite{bul01}):
\begin{equation}
\label{eq:vint}
 \log \left(\frac{v_{\rm int}}{\rm km\: s^{-1}}\right) 
= 2.09+0.0404 z_f 
+0.29 \left[\log\left(\frac{M_{\rm gal} h}
{M_{\odot}}\right)-11.42\right]\;,
\end{equation}
where $M_{\rm gal}$ is the mass of the galaxy and $h$ is the Hubble
constant represented by $H_0=100 h\;\rm km\: s^{-1}\:
Mpc^{-1}$. Equation~(\ref{eq:vint}) can also be applied to clusters. We
compare the VDFs with the observed ones. Here, `internal velocity' means
rotation velocity (late type galaxies) or dispersion velocity (early
type galaxies). More massive galaxies have larger internal
velocities. Moreover, there is a benefit to use VDFs instead of
MDFs. The tidal interaction among galaxies in a cluster and that between
a galaxy and the host cluster somewhat affect the MDFs and VDFs after
their evolutions froze \citep{kly99,oka99,ghi00}. This makes it
difficult to estimate the time of the freeze from the MDFs or VDFs
because the SPS model does not treat the interactions. However, recent
ultra-high resolution N-body simulations showed that the VDFs are much
less affected by the tidal interactions than the MDFs
\citep{kly99,oka99,ghi00}. We have confirmed that the VDFs derived using
the SPS model are consistent with those derived by several simulations
\citep{fuj02a}.

To compare the theoretical model with observations, we use the
catalogues of early (E+S0) and late (Sp) type galaxies in seven nearby
clusters \citep{gio97a,sco98b}. The clusters are A1367, Coma (A1656),
A2634, A262, Pegasus and Cancer clusters and the NGC~507 group, although
we do not discuss the Cancer cluster because the number of galaxies is
too small (less than 10 within the cluster virial radius). We do not
discriminate the NGC~383 group from the NGC~507 group because they are
very close to each other. Cluster membership criteria are described in
section~8 of \citet{gio97a}. In the SPS model, we fix the masses of the
clusters at those obtained from X-ray observations if available
\citep{rei02}. The X-ray mass estimation does not suffer from the small
number of galaxies especially for small clusters. Reiprich and B{\"
o}hringer (2002) assumed that the average density of a cluster is 200
times the critical density of the universe. This is appropriate for the
Einstein de-Sitter universe but not for the low density universe we
adopted ($\Omega_0=0.3$, $\lambda=0.7$, $\sigma_8=1.0$, and $h=0.7$). In
this paper, we assume that the average density of a cluster is 100 times
the critical density of the universe, and the cluster masses in Reiprich
and B{\" o}hringer (2002) are modified to be consistent with the
assumption. On the modification, we assume that the dark matter
distribution in clusters is approximated by a power law distribution
with an index of $-2.4$ \citep{nav97,hor99}. For the Pegasus cluster,
since the mass is not obtained by X-ray observations, we estimate it by
a relation between the velocity dispersion of galaxies in a cluster and
the mass of the cluster; we assume that the relation is also given by
equation (\ref{eq:vint}) for $z_f = 0$. The velocity dispersion of the
Pegasus cluster is $393\rm\; km\; s^{-1}$ \citep{sco98b}. The resultant
cluster masses ($M_{\rm cl}$) and viral radii ($r_{\rm vir}$) are shown
in table~\ref{tab1}. Note that these mass estimations assume that
clusters and groups have collapsed and virialized. On the other hand, we
leave the average redshift when galaxies stopped growing in mass ($z_f$)
to be a free parameter in the SPS model. In order to deal with the
observed rotation velocities of late type galaxies and the observed
dispersion velocities of early type galaxies at the same time, we
multiply the dispersion velocities by 1.5 to obtain the effective
rotation velocities of the early type galaxies, because the rotation
velocity of a late type galaxies ($v_{\rm rot}$) and the dispersion
velocity of an early type galaxy ($\sigma$) are respectively given by
\begin{equation}
 v_{\rm rot}^2 \approx \frac{G M_{\rm gal}}{r_{\rm gal}}\:, \:
 \sigma^2 \approx \frac{1}{\alpha}\frac{G M_{\rm gal}}{r_{\rm gal}}\:,
\end{equation}
where $G$ is the gravitational constant, $\alpha$ is the constant ($\sim
2-3$), and $r_{\rm gal}$ is the radius of the galaxy. Thus, we define
internal velocities as $v_{\rm int}=v_{\rm rot}$ for late type galaxies
and $v_{\rm int}=1.5\sigma$ for early type galaxies. We assume that each
dark halo contains only one galaxy. 
% ## changed
Actually, N-body numerical simulations show that a dark halo contained
in a cluster does not contain smaller dark halos (subhalos in a subhalo)
at least for dark halos of galaxy scales (Fig.~4 in \cite{spr01}).  This
is also confirmed in calculations including an N-body simulation and
semi-analytic model of galaxy formation (Okamoto 2002, private
communications). Mostly, a dark halo of galaxy scales has only one
galaxy.
% ## end
One may think that this is because of a poor numerical
resolution. However, the time scale of dynamical friction for a subhalo
in a galactic scale dark halo is
\begin{equation}
 t_{\rm fric}= 3.3\times 10^{9} \left(\frac{\ln \Lambda}{3}\right)^{-1}
\left(\frac{r_{\rm dh}}{60\rm\: kpc}\right)^2
\left(\frac{v_{\rm int}}{220\rm\: km\: s^{-1}}\right)
\left(\frac{M_{\rm sdh}}{2\times 10^{10}\: M_{\odot}}\right)
^{-1}\:\rm yr\;,
\end{equation}
where $\ln \Lambda$ is the Coulomb logarithm, $r_{\rm dh}$ is the radius
of the galactic scale dark halo, $v_{\rm int}$ is the internal velocity
of the galactic scale dark halo, and $M_{\rm sdh}$ is the mass of a
subhalo in the galactic scale dark halo \citep{bin87}. This means that
for a galactic scale dark halo, most subhalos in it should have merged
with the galaxy at the center of the dark halo within the time scale of
cluster evolution ($\sim 10^{10}$~yr) except for very small subhalos
that would be difficult to be observed. We do not discuss dark halos of
group scales ($\gtrsim 5\times 10^{12}\: M_{\odot}$) in clusters unless
otherwise mentioned; those dark halos may contain multiple galaxies.
% ## add
The number of galaxies included in those halos is small (see
\S\ref{sec:results}).

\section{Results}
\label{sec:results}

\begin{figure}
  \begin{center}
    \FigureFile(120mm,120mm){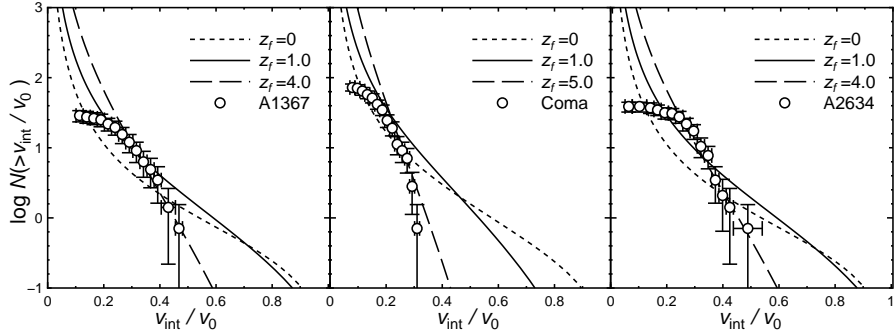}
    %%% \FigureFile(width,height){filename}
  \end{center}
  \caption{Galaxy VDFs for $z_f=0$ (dotted lines), $z_f=1.0$ (solid
lines), and $z_f=4.0$ or~5.0 (dashed lines) for A1367, Coma, and A2634
clusters. The VDFs obtained by observations are shown by
circles.}\label{fig:cluster}
\end{figure}

In Figure~\ref{fig:cluster}, we present the predicted and observed VDFs
for A1367, Coma, and A2634 clusters. These clusters are relatively
massive ($\gtrsim 4\times 10^{14}\; M_{\odot}$; table~\ref{tab1}). The
VDFs are represented by cumulative numbers, that is, the numbers of
galaxies having internal velocities larger than a certain value. The
internal velocities are normalized by the velocity dispersions of
galaxies in the host clusters ($V_0$; table~\ref{tab1}), which are
converted from the cluster masses using equation~(\ref{eq:vint}) for
$z_f=0$ for consistency with galaxy internal velocities. The
observational data are complete for $v_{\rm int}\gtrsim 200\rm\: km\:
s^{-1}$ \citep{gio97b,sco97}. With the observational data, we count the
two-dimensional numbers of galaxies on the celestial sphere within the
angles corresponding to the cluster virial radii ($N_{\rm 2D}$;
table~\ref{tab1}). Then, we multiply the number by 0.7 to exclude
projected galaxies within the angles but outside the cluster virial
radii in the three-dimensional space. The number 0.7 is appropriate when
the galaxy distribution in clusters is approximated by a power law
distribution with an index of $-2.4$ and has a cut off at several times
the virial radii. The index of the power low distribution is consistent
with observations and the results of numerical simulations if galaxy
distribution follows dark matter distribution in clusters
\citep{nav97,hor99}. Most of the galaxies left in our samples ($\gtrsim
70$~\%), especially for massive clusters, are early type galaxies that
are mainly found in the central regions of
clusters. Figure~\ref{fig:cluster} shows that the VDFs calculated using
the SPS model can reproduce the observations if the galaxies in those
clusters formed at redshifts of about 4. The deviation of the
observational data from the predictions at small velocities is due to
the incompleteness of the observations of small galaxies. For the Coma
cluster, the largest velocity point also deviates from the line. This
can be attributed to the dynamical friction that forces massive galaxies
to fall into the cluster center and to merge with a central cD galaxy
\citep{fuj02a}. The derived formation redshift of the cluster galaxies
is comparable to the redshift of major star formation of early type
galaxies in massive clusters \citep{kod98}. Since most of the galaxies
in our samples are early type galaxies, this explicitly shows that early
type galaxies in massive clusters are very old and have not grown for a
long time in terms of their masses as well as the stars in them. Mass
growth of the early type galaxies (e.g. mergers between galaxies) may be
essential for the star formation.  We note that if we adopt the
Einstein-de Sitter universe as cosmological parameters ($\Omega_0=1$,
$\lambda=0$, $\sigma_8=0.7$, and $h=0.5$), smaller formation redshifts
($z_f\sim 2$) are preferable to be consistent with the observations.

\begin{figure}
  \begin{center}
    \FigureFile(120mm,120mm){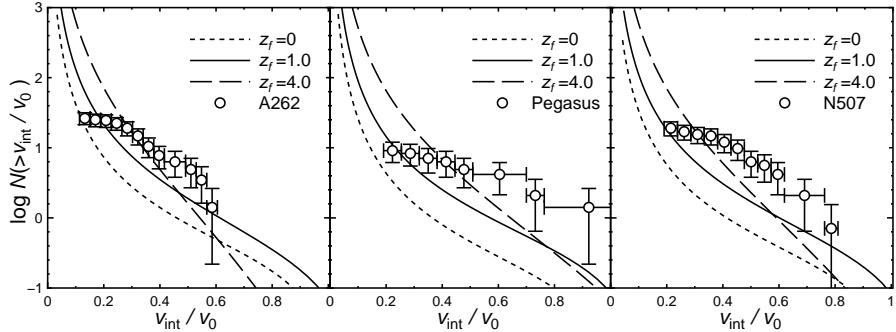}
    %%% \FigureFile(width,height){filename}
  \end{center}
  \caption{Same as Figure~\ref{fig:cluster} but for A262,
the Pegasus cluster, and the NGC~507 group.}\label{fig:group}
\end{figure}

Figure~\ref{fig:group} shows the VDFs for A262, the Pegasus cluster, and
the NGC 507 group. These clusters have relatively small masses
($\lesssim 1\times 10^{14}\; M_{\odot}$; table~\ref{tab1}) and we here
call them groups to discriminate them from the above massive
clusters. For a given $V_{\rm int}/V_0$, the mass of a subhalo in the
groups is smaller than that in the massive clusters. In the hierarchical
clustering scenario, less massive halos form earlier. This means that
for a given redshift ($z\geq z_f$), hierarchical clustering proceeds
further on the mass scale of group subhalos than on the mass scale of
cluster subhalos. Thus, for a large $V_{\rm int}/V_0$ and $z_f$, the
predicted cumulative numbers $N(>V_{\rm int}/V_0)$ of the groups are
larger than those of the massive clusters (Figures~\ref{fig:cluster}
and~\ref{fig:group}).

\begin{figure}
  \begin{center}
    \FigureFile(120mm,120mm){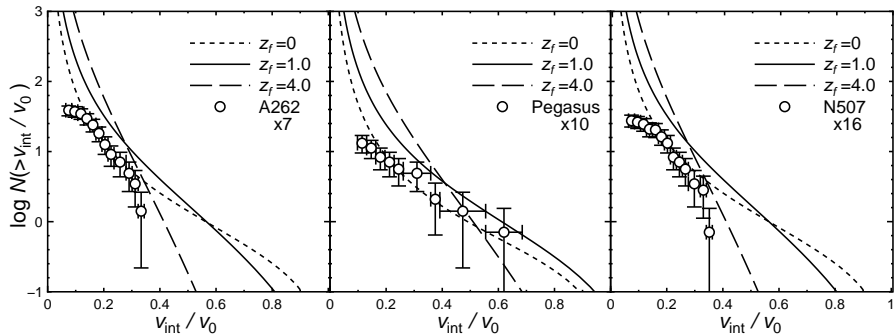}
    %%% \FigureFile(width,height){filename}
  \end{center}
  \caption{Same as Figure~\ref{fig:group}
but the group masses are increased so that the model predictions match
the observations.}\label{fig:infant}
\end{figure}

Figure~\ref{fig:group} shows that contrary to A1367, Coma, and A2634,
the observed VDFs of the groups do not match the SPS model predictions
regardless of $z_f$; these groups have too many massive galaxies for
their masses. For late type galaxies, \citet{gio97b} show that the
observational data are complete for $v_{\rm int}\gtrsim 200\rm\: km\:
s^{-1}$ (A262 and NGC~507) and $\gtrsim 130\rm\: km\: s^{-1}$
(Pegasus). For early type galaxies, although the completeness of the
data is not commented in the catalogue, we confirm that the data are
complete at least to the completeness limit velocity of the late type
galaxies from the magnitude distribution. The excess of observed
galaxies in Figure~\ref{fig:group} suggests that the actual masses of
these groups are much larger than their apparent masses estimated on the
assumption that they are relaxed and virialized. For these groups, it
may indicate that matter including galaxies has already been gathered
through gravity, but it has not virialized and relaxed to be matured
massive clusters. That is, they may be `infant galaxy clusters' just
before virialization, although it is one of possible
solutions. Figure~\ref{fig:infant} actually supports this
interpretation. This figure shows that the theoretical predictions match
the observations perfectly if the group masses are larger than their
apparent masses; the masses are assumed to be 7 times (A262), 10 times
(Pegasus), and 16 times (NGC 507) larger than their apparent masses. The
group radii and velocity dispersions are increased with the group masses
(table~\ref{tab1}). Strictly speaking, if these groups have not fully
relaxed, it is not exact to estimate the galaxy numbers within the
virial radii by multiplying the observed two-dimensional numbers by 0.7
as is the case where groups are relaxed. However, at least this
estimation and the comparison with the SPS model predictions should be
appropriate as long as the index of the galaxy distribution is not much
different from $-2.4$. In table~\ref{tab1}, $N_{\rm 2D}$ of each group
increases as its $M_{\rm cl}$ and $r_{\rm vir}$ increase. If we assume
that the galaxy density distribution is represented by $r^{-2.4}$, the
number of galaxies within a radius $r$ has the relation of $N \propto
r^{0.6}$. Since the projection of outskirt galaxies does not make
significant difference between $N$ and $N_{\rm 2D}$, the increase of
$N_{\rm 2D}$ is roughly consistent with this prediction. We adopt the
lines of $z_f = 0$ to be compared with the observational data because
these groups are growing at present and thus $z_f$ is expected to be
close to zero. In fact, while the lines of $z_f\sim 1$ can also be
adjusted to match the observations, the lines of a higher redshift
($z_f\sim 4$) cannot because the slopes of these lines are different
from the slopes of the observational data. This actually indicates that
the galaxies formed recently in these group regions. In
Figure~\ref{fig:infant}, the deviation of the observational data from
the predictions at small and large velocities can be explained by the
incompleteness of observations of small galaxies and dynamical friction,
respectively.

% ## add
We make a comment on subhalos in a subhalo here. Our model does not take
account of galaxies in larger subhalos of galaxy group scales. However,
we can show that the number of those galaxies is small as follows. We
take A1367 as an example. The expected number of subhalos with the
scales of galaxy groups (say $V_{\rm int} \sim 400\rm\: km\: s^{-1}$) in
this cluster is only about one (Figure~\ref{fig:cluster}). On the other
hand, the internal velocity $V_{\rm int}$ of the subhalo is close to
that of the Pegasus cluster ($V_0$; table~\ref{tab1}). As is shown by
the lines in Figure~\ref{fig:group}, the expected number of galaxies in
a (sub)halo of this mass is $\sim 1/10$ of the expected number of
galaxies in A1367 for a given $V_{\rm int}$. This means that the
galaxies in larger subhalos do not much affect our results shown above.

\section{Discussion}

If the groups we investigated are infant galaxy clusters, we expect that
they will relax and virialize within their dynamical time-scales ($\sim
10^9$~yrs). Considering the selection effect that groups having many
galaxies in small regions can easily be observed, many of groups
observed so far may be such infant galaxy clusters. In particular,
compact galaxy groups such as the Hickson groups \citep{hic82}, which
are small galaxy groups that are crowded with galaxies, are good
candidates of infant galaxy clusters. Some of these groups may be
collapsing and virializing at present. In fact, observations support the
idea. First, the properties of compact groups (e.g. the X-ray
luminosities or the spatial distribution of the galaxies) vary widely
among the groups \citep{pil95,rib98}. This may reflect that we are
seeing different dynamical stages of cluster or group evolution. Second,
HCG 62, which is one of the Hickson groups, has hard X-ray emission
\citep{fuk01}. This may be the result of gas accretion toward the group
and the particle acceleration at the shock waves or in turbulence formed
in the rapidly accreting gas; the accelerated particles could produce
the nonthermal hard X-ray emission via inverse Compton scattering of
photons \citep{fuj01,fuj02b}. In the future, direct observations of
matter distribution around groups through gravitational lensing will be
particularly important to reveal the process of cluster formation. For
example, diffuse density enhancement may be observed around some
groups. It will also be important to compare the results of the SPS
model with those of gravitational lensing, optical, and X-ray
observations for many clusters and groups systematically to reveal the
whole picture of cluster formation.

Finally, we note that \citet{evr96} used cosmological gasdynamic
simulations to investigate the accuracy of cluster mass estimates based
on X-ray observations; they found that X-ray mass estimates are very
accurate and the errors are 14\%-29\%. However, recent X-ray
observations have shown that the structure of clusters is much more
complicated than the results of numerical simulations done by
\citet{evr96} and have shown that many clusters have not fully
relaxed. \citet{ike96} found that the Fornax cluster has two distinct
spatial scales of the gravitational potential. Recently, Chandra
observatory found that clusters often have substructures with small
X-ray temperatures called `cold fronts' \citep{mar00,vik01,maz01}. These
show that clusters often have dark halos with the scales of galaxy
groups, which is also shown by recent ultra-high resolution simulations
\citep{oka99,ghi00,spr01}. For the three groups we investigated, the
observations of X-ray temperature (or velocity dispersion of galaxies)
might have been made only for possible central substructures, because it
is often difficult to observe the X-ray temperatures (or the velocity
dispersion of galaxies) of the relatively diffuse outer regions of the
possible host clusters, especially when the host clusters are
small. This could lead to the underestimation of the host cluster
masses. The hard X-ray emission from HCG 62 may come from the diffuse
hotter gas in the outer region of the cluster. If this kind of emission
is observed in groups with an excess of galaxies, it would be an evidence
that the groups are infant clusters.

\section{Conclusions}

We study the velocity distribution functions (VDFs) of galaxies in
clusters. We show that galaxies in some clusters stopped growing at a
very early time of the universe or redshift of $\sim 4$, which is
comparable to the redshift of major star formation in early type
galaxies in clusters. Moreover, we find that some galaxy groups have too
many massive galaxies for their apparent masses. One possibility is that
these groups are in a transient phase and are much more massive than
their apparent masses. If this is the case, they should be called
`infant galaxy clusters' that will be matured clusters in the dynamical
time-scale ($\sim 10^9$ yrs). Considering the selection effect for
observations, many of galaxy groups observed so far may be such infant
clusters.

\vspace{10mm}

 I thank C.~L. Sarazin, H. Yahagi, M. Nagashima, T. Kodama, S. Nishiura,
 and T. Okamoto for discussions. Y.~F.\ was supported in part by a
 Grant-in-Aid from the Ministry of Education, Science, Sports, and
 Culture of Japan (14740175).

%%%%%%%%%%%%%%%%%%%%%%%%%%%%%%%%%%%%%%%
\newpage

\begin{table}
  \caption{Parameters for Clusters.}\label{tab1}
  \begin{center}
    \begin{tabular}{ccccc}
  \hline\hline
      Cluster & $M_{\rm cl}$ & $r_{\rm vir}$ & $V_0$ & $N_{\rm 2D}$ \\
              & ($10^{14}\; M_{\odot}$) & (Mpc) & ($\rm km\: s^{-1}$)& \\
  \hline
      A1367  & 4.8  & 2.0 & 988  & 41  \\
      Coma   & 16   & 3.0 & 1410 & 103 \\
      A2634  & 4.5  & 2.0 & 971  & 56  \\
      A262   & 1.2  & 1.6 & 661  & 38 \\
      A262$^\dagger$ ($\times 7$)   
             & 8.4  & 2.4  & 1163 & 55 \\
      Pegasus& 0.20 & 0.69 & 393 & 13 \\
      Pegasus$^\dagger$ ($\times 10$)
             & 2.0  & 1.5 & 766 & 19 \\
      NGC507 & 0.54 & 0.96 & 525 & 27 \\
      NGC507$^\dagger$ ($\times 16$) 
             & 8.6 & 2.4 & 1173 & 39 \\
  \hline
    \end{tabular}
  \end{center}
$\dagger$ The cases where the masses are increased.
\end{table}

\end{document}